\documentclass[twocolumn,aps,epsfig,nofootinbib]{revtex4}

\usepackage{graphicx}
\usepackage{epstopdf}
\usepackage{latexsym}
\usepackage{amssymb}
\usepackage{amsmath}
\usepackage{color}
\usepackage{mathrsfs}

\usepackage[center]{subfigure}

\begin{document}

 \newcommand{\bq}{\begin{equation}}
 \newcommand{\eq}{\end{equation}}
 \newcommand{\bqn}{\begin{eqnarray}}
 \newcommand{\eqn}{\end{eqnarray}}
 \newcommand{\nb}{\nonumber}
 \newcommand{\lb}{\label}
\newcommand{\PRL}{Phys. Rev. Lett.}
\newcommand{\PL}{Phys. Lett.}
\newcommand{\PR}{Phys. Rev.}
\newcommand{\PRD}{Phys. Rev. D.}
\newcommand{\CQG}{Class. Quantum Grav.}
\newcommand{\JCAP}{J. Cosmol. Astropart. Phys.}
\newcommand{\JHEP}{J. High. Energy. Phys.}

 \title{Hawking radiation of charged Einstein-aether black holes at both Killing and universal horizons}

\author{Chikun Ding${}^{a, b}$ }
\email{Chikun_Ding@baylor.edu}
\author{ Anzhong Wang$^{b, c}$\footnote{Corresponding author}}
 \email{Anzhong_Wang@baylor.edu}
 \author{Xinwen Wang${}^{b,c,d}$ }
 \email{Xinwen_Wang@baylor.edu}
\author{Tao Zhu${}^{b, c}$ }
\email{Tao_Zhu@baylor.edu}
  \affiliation{$^{a}$ Department of Information Science and Technology, Hunan University of Humanities, Science and Technology, Loudi, Hunan
417000, P. R. China\\
$^{b}$ GCAP-CASPER, Physics Department, Baylor University, Waco, Texas 76798-7316, USA\\
$^{c}$ Institute for Advanced Physics \& Mathematics, Zhejiang University of Technology, Hangzhou, 310032, China\\
$^{d}$Departamento de F\'{\i}sica Te\'orica, Instituto de F\'{\i}sica, UERJ, 20550-900, Rio de Janeiro, Brazil}

\date{\today}

\begin{abstract}

We study analytically quantum tunneling of  relativistic and non-relativistic particles  at both  Killing and universal horizons of  Einstein-Maxwell-aether
black holes, after high-order curvature corrections are taken into account, for which the dispersion relation of the particles becomes nonlinear.  Our
results at the Killing horizons confirm the previous ones, i.e., at high frequencies the corresponding radiation remains thermal and the nonlinearity
of the dispersion does not alter the  Hawking radiation significantly. In contrary, non-relativistic particles are  created at universal horizons and are
radiated out to infinity. The radiation also has a thermal spectrum, and the corresponding  temperature takes the form, $T^{z}_{UH} = 2\kappa_{UH}
(z-1)/(2\pi z)$, where $z$  denotes the power of the leading term in the nonlinear dispersion relation,  $\kappa_{UH}$ is the surface gravity of the 
universal horizon, defined by peering behavior of ray trajectories at the universal horizon. We also study the Smarr formula by assuming that: (a) the 
entropy is proportional to the area of the universal horizon, and (b) the first law of black hole thermodynamics   holds, whereby we derive the Smarr 
mass, which in general is different from the  total  mass obtained at infinity.  This indicates that one or both of these assumptions must be modified.

\end{abstract}


\maketitle

\section{Introduction}
\renewcommand{\theequation}{1.\arabic{equation}} \setcounter{equation}{0}

In the Einstein-aether theory, a timelike aether vector field is introduced to describe extra degrees of the gravitational sector,  in addition to the spin-2 ones found
in general relativity that move with the speed of light  \cite{jacobson3}. In fact, due to the presence of the aether field, spin-0 and spin-1 particles are also present,
and  all  move  at different speeds \cite{jacobson4}. Moreover, due to Cherenkov effects they  must move with speeds no less than that of light \cite{EMS}. It should
be noted  that  here the propagations faster than that of the light do not violate causality \cite{jacobson3}. In particular, gravitational theories with breaking Lorentz invariance
(LI) still allow the existence of black holes \cite{BS11,barausse,berglund,lin,UHs,bhattacharyya3}. However, instead of Killing horizons, now the boundaries of black
holes are  hypersurfaces, termed as {\em universal horizons}, which are always inside Killing horizons and trap excitations traveling at arbitrarily high velocities. {
The crucial ingredient for the existence of a universal horizon is  the presence of a globally timelike foliation of the spacetime \cite{BS11}.
Such a preferred foliation, for example,  naturally rises in the Horava theory \cite{Horava}. But in the Einstein-aether theory this is true only when the aether is
hypersurface-orthorgonal  \cite{bhattacharyya3,Jacobson10}. This is always the case  in spherically symmetric spacetimes, although   in other spacetimes, such as the ones
with rotation, the aether is generically not hypersurface-orthorgonal \cite{bhattacharyya3,Jacobson10}. With the above in mind,} a
  slightly modified first law of black hole mechanics was found to exist for the neutral Einstein-aether black holes \cite{berglund}, but for the charged  Einstein-aether
black holes, such a  law  is still absent  \cite{ding}.

Berglund {\it et al} \cite{berglund2} used tunneling method to study the corresponding Hawking radiation at the universal horizon for a scalar field that violates the local LI,
and found that the universal horizon radiates as a blackbody at a fixed temperature. Using a collapsing null shell, on the other hand, Michel and Parentani  \cite{michel}
computed the late time radiation and found  that the mode pasting across the shell is adiabatic at late time. This implies that large black holes emit a thermal flux with
a temperature fixed by the surface gravity of the Killing horizon. This, in turn, suggests that the universal horizon should play no role in the thermodynamical properties
of these black holes. However, it should be noted that in such a setting, the khronon field is not continuous across the collapsing null shell \cite{NoteA}. Normally, it is
expected that such discontinuities should not affect the final results \cite{michel}. However,  the khronon field here plays a special role, and in particular
it defines the causality of the spacetime. So far, it is not clear whether the results presented in \cite{michel} will remain the same or not, after the continuity of the aether field
 across the collapsing surface is assumed.

Another different approach was taken by Cropp {\it et al}  \cite{cropp}, in which ray trajectories in such black hole backgrounds were studied, and  evidence was found, which  shows that
Hawking radiation is associated with the universal horizon, while the ``lingering" of low-energy ray trajectories near the Killing horizon hints a reprocessing there.

 In this paper, we have no intention to resolve the above discrepancy, but rather study the Hawking radiation at both  universal and  Killing horizons of the charged
 Einstein-aether black holes found in \cite{ding}. Although we  also use the tunneling approach, we shall give up the null geodesic method \cite{parikh}. Instead, we shall adopt the
 Hamilton-Jacobi  method \cite{agheben,acquaviva,ding1,ding2}, and show that particles with $z \ge 2$ are indeed created at the universal horizon, and the  corresponding Hawking radiation   is  thermal, where $z$
 characterizes the nonlinearity of the  dispersion relation,  appearing in Eq.(\ref{ksquare}) given below. Although for any given $z \ge 2$ the universal horizon radiates thermally,  particles with different $z$
 will feel different temperatures, given by
 \bq
 T^z_{UH} = \left(2 - \frac{2}{z}\right) \frac{\kappa_{UH}}{2\pi},
 \eq
 where $\kappa_{UH}$ is the surface gravity of the universal horizon, defined by peering behavior of ray trajectories at the
universal horizon \cite{cropp,lin,ding}.   On the other hand,  in high frequencies only relativistic particles  are created at the Killing horizon, and the
 corresponding Hawking radiation is  the same as that obtained in general relativity \cite{Carlip}. This is consistent with previous findings \cite{unruh} 
 \footnote{It should be noted that in low frequencies the Hawking radiation
 is sensitive to high-order corrections. For detail, see, for example,  \cite{MP09}.}. To confirm further this result, we use a completely different approach 
 --- the extended WKB method (the Bremmer series)  \cite{countant12,coutant16}, and 
  reach the same conclusion.

Specifically,  the paper is organized as follows. In Sec. II we give a brief review of the Einstein-aether theory and the charged black holes obtained in \cite{ding}, while in Sec. III we study the tunneling of
spin-0 particles with a nonlinear dispersion. In Sec. IV,  we study the Smarr formula by assuming that the first law of black hole mechanics holds at the universal horizon,
 and find the corresponding Smarr mass, which in general is quite different from    the Arnowitt-Deser-Misner (ADM) mass at infinity.  In Sec. V, we present our main conclusions.
 Two appendices are also included. In Appendix A,   some useful formulas  for fractional derivatives are presented, while in Appendix B, we study the Hawking radiation 
 at the Killing horizon for particles with the $z = 2$ nonlinear dispersion relation,
 by using the extended WKB method (the Bremmer series), and obtain the same results obtained in Section III by using the Hamilton-Jacobi  method, as it is expected.

\section{Einstein-Maxwell-aether theory and charged black holes}
\renewcommand{\theequation}{2.\arabic{equation}} \setcounter{equation}{0}

The  Einstein-Maxwell-aether theory considered in \cite{ding} is described by the action,
\begin{eqnarray}\label{lag}
\mathcal{S}=
\int d^4x\sqrt{-g}\Big[\frac{1}{16\pi G_{\ae}}(\mathcal{R}+\mathcal{L}_{\ae})+\mathcal{L}_M\Big], \label{action}
\end{eqnarray}
where  $G_{\ae}$ is a coupling constant of the theory, and is related to Newton's gravitational constant $G_{N}$ by $G_{\ae}=(1-c_{14}/2)G_N$ \cite{eling}.
$\mathcal{R}$ is the four-dimensional (4D) Ricci scalar,   $\mathcal{L}_M$ denotes the matter Lagrangian density, and
$\mathcal{L}_{\ae}$ the aether Lagrangian density, defined as
\begin{eqnarray}
-\mathcal{L}_{\ae}=Z^{ab}_{~~cd}(\nabla_au^c)(\nabla_bu^d)-\lambda(u^2+1),
\end{eqnarray}
where $\nabla_{\mu}$ denotes the covariant derivative with respect to the 4D metric $g_{ab}$, which has the signatures ($-, +, +, +$).  $u_{a}$ is the four-velocity of the aether,
$\lambda$  a Lagrangian multiplier that guarantees $u_{a}$ to be timelike, and $Z^{ab}_{~~cd}$ is defined as \cite{eling,garfinkle},
\begin{eqnarray}
Z^{ab}_{~~cd}=c_1g^{ab}g_{cd}+c_2\delta^a_{~c}\delta^b_{~d}
+c_3\delta^a_{~d}\delta^b_{~c}-c_4u^au^bg_{cd}, ~~~
\end{eqnarray}
where $c_i$'s  are coupling constants of the theory. There are a number of theoretical and observational bounds on the coupling constants $c_i$ \cite{jacobson5}. Here,  we impose the following
constraints \cite{ding}, $0\leq c_{14}<2,\; 2+c_{13}+3c_2>0,\; 0\leq c_{13}<1$,
where $c_{14}\equiv c_1+c_4$, and so on.
The source-free Maxwell Lagrangian $\mathcal{L}_M$ is given by
\begin{equation}
\mathcal{L}_M=-\frac{1}{16\pi G_{\ae}}\mathcal{F}_{ab}\mathcal{F}^{ab},
~\mathcal{F}_{ab}=\nabla_a\mathcal{A}_b-\nabla_b\mathcal{A}_a,
\end{equation}
where $\mathcal{A}_a$ is the  four-vector of the electromagnetic field.

The static spherically symmetric spacetimes in    the Eddington-Finklestein coordinates are described by the metric \cite{JB13},
\begin{eqnarray}\label{metric}
ds^2=-e(r)dv^2+2dvdr+r^2d\Omega^2,
\end{eqnarray}
where $d\Omega^2 \equiv d\theta^2 + \sin^2\theta d\phi^2$. The corresponding time-translation Killing  and aether vectors are given, respectively, by
\bqn
&&
  \chi^a=\delta^{a}_{v},\quad u^a= \alpha \delta^{a}_{v}  + \beta \delta^{a}_{r},
\eqn
where $\alpha,\;\beta$ are functions of $r$ only, and the constrain is $u^2=-1$.
Introducing the spacelike unit vector $s_a$ via the relations $u^as_a=0,\;s^2=1$,
we find that  the metric can be written as
\bq
\lb{tetrad}
g_{ab}=-u_au_b+s_as_b+\hat{g}_{ab},
\eq
where $\hat{g}_{ab} \equiv {\mbox{diag}}\left(0, 0, r^2, r^2\sin\theta^2\right)$, and  that,
\bqn
\label{abe}
 && \alpha(r)=\frac{1}{(s\cdot\chi)-(u\cdot\chi)},\quad \beta(r)=-(s\cdot\chi), \nb\\
 &&  e(r)=(u\cdot\chi)^2-(s\cdot\chi)^2.
\eqn

The Killing horizon is the location where $\chi^a$ becomes null, i.e., $e(r_{KH}) = 0$.

The universal horizon, on the other hand,  is the located at $(u\cdot \chi) = 0$ \cite{BS11,barausse},  that is,
 \bq
 \lb{uh}
\left. \left(e\alpha^2 + 1\right)\right|_{{UH}} = 0.
 \eq
 The surface gravity at the universal horizon is defined as \cite{cropp},
 \bqn
 \lb{surfaceG}
 \kappa_{UH} &\equiv& \left. \frac{1}{2}\nabla_{u}\left(u\cdot\chi\right)\right|_{UH} \nb\\
 &=&\left. \frac{1}{2}\left(a\cdot s\right)\left(s\cdot\chi\right)\right|_{UH},
  \eqn
 which is precisely the one  obtained from the peeling behavior of rays propagating with infinite group velocity with respect to the aether as shown explicitly in \cite{lin,cropp}.

In \cite{ding}, two classes of the charged Einstein-aether black hole solutions were found in closed forms, for particular choices of the coupling constants $c_{i}$'s.
 They are given as follows.

\subsection{Exact Charged Einstein-aether Solutions for $c_{14} = 0$}

When $c_{14} = 0$, which corresponds to the case in which the spin-0 particle of the khronon field has an infinitely large velocity,  the charged Einstein-aether black hole solutions are
given by  \cite{ding},
\begin{eqnarray}\label{aether}
 &&(s\cdot\chi)=\frac{r^2_{\ae}}{r^2},\nonumber\\
 &&(u\cdot\chi)=-\sqrt{1-\frac{r_0}{r}+\frac{Q^2}{r^2}+\frac{(1-c_{13})
 r^4_{\ae}}{r^4}},\nb\\
&& e(r)=1-\frac{r_0}{r}+\frac{Q^2}{r^2}-\frac{c_{13}r^4_{\ae}}{r^4},
\end{eqnarray}
where $r_0, r_{\ae}$ and $Q$ are the integration constants, and $Q$ is related to the Maxwell field via the relation,
 \bq
 \lb{em}
 {\cal{F}}_{ab} = \frac{Q}{r^2}(u_as_b - u_b s_a).
 \eq
In order for the khronon field to be well-defined in the whole spacetime, the integration constant $r_{\ae}$ must be given by \cite{ding},
 \bq
 \lb{r4}
 r^4_{\ae}=\frac{1}{1-c_{13}}\left(r_{UH}^4-\frac{1}{2}r_0r_{UH}^3\right),
\eq
where $r_{UH}$ is the location of the universal horizon, given by
\bq
\lb{ruh1}
r_{UH}=\frac{r_0}{2}\left(\frac{3}{4}+\sqrt{\frac{9}{16}-2\frac{Q^2}{r_0^2}}\right).
\eq
The location of the Killing horizon is at $r = r_{KH}$, given by,
\bqn
\lb{rkh1}
 r_{KH}=\frac{r_0}{2}\left(\frac{1}{2}+L
 +\sqrt{N-P+\frac{1-4Q^2/r_0^2}{4L}}\right),  ~~~
 \eqn
 where
 \bqn
 \lb{rkh1b}
 && L=\sqrt{\frac{N}{2}+P}, \quad N=\frac{1}{2}-\frac{4Q^2}{3r_0^2},\nonumber\\
 && P=\frac{2^{1/3}(12I+Q^4/r_0^4)}{3H}+\frac{H}{3\cdot2^{1/3}},\nb\\
&&  I=-\frac{c_{13}}{1-c_{13}}\left(\frac{r_{UH}^4}{r_0^4}
 -\frac{r_{UH}^3}{2r_0^3}\right),\nonumber\\
 &&
 H=\left(J+\sqrt{-4(12I+Q^4/r_0^4)^3+J^2}\right)^{1/3},\nb\\
 &&  J=27I-72IQ^2/r_0^2
 +2Q^6/r_0^6.
\end{eqnarray}

\subsection{Exact Charged Einstein-aether Solutions for $c_{123} = 0$}

When $c_{123} = 0$, the velocity of the spin-0 particle of the khronon field is zero, and the solutions are given by,
\bqn
\lb{classB}
 &&(u\cdot\chi)=-1+\frac{r_0}{2r},\;\;\; (s\cdot\chi)=\frac{r_0+2r_u}{2r},\nb\\
 && e(r)=1-\frac{r_0}{r}-\frac{r_u(r_0+r_u)}{r^2},
 \eqn
 where $r_0$ is a non-negative  integration constant, and $r_u$ is given by,
 \bq
  r_u=\frac{r_0}{2}\left(\sqrt{\frac{\mathrm{p}}{\mathrm{g}}-\frac{4Q^2}
 {\mathrm{g}r_0^2}}-1\right),
 \eq
where
\bq
\mathrm{g}\equiv1-c_{13},\;\mathrm{p}\equiv1-\frac{c_{14}}{2}.
\eq
The locations of the universal and Killing horizons are given, respectively, by
 \bqn
 \lb{ruh_rkh2}
  r_{UH}=\frac{r_0}{2},\quad  r_{KH}=r_0+r_u.
 \eqn
 It should be noted that, in order to have the khronon field well-defined in the whole spacetime,
  in the present case we must assume that
  \bq
 |Q| \leq\frac{1}{2}\sqrt{\mathrm{p}-\mathrm{g}}\;r_0, \quad
\mathrm{p}\geq\mathrm{g}.
\eq

\section{ Hawking Radiation with Nonlinear Dispersion Relation}
\renewcommand{\theequation}{3.\arabic{equation}} \setcounter{equation}{0}

The semi-classical tunneling approximations that model the Hawking radiation usually follow two approaches, { the null geodesics (NG) method} explored
by Parikh and Wilczek  \cite{parikh},  and { the Hamilton-Jacobi (HJ) method} used by Agheben {\it et al} \cite{agheben,ding1,ding2,acquaviva}. { Since the final
results should not depend on the methods to be used, in this paper we choose the HJ method. } In each method, particles with positive (negative) energy
 just inside (outside) of the horizon are assumed to escape (fall into) it. Both of the processes are forbidden classically, so the radiation is quantum mechanical
 in nature.

In the semi-classical approximation, the charged massless scalar field $\phi(x)$ can be written as $\phi(x) =\phi_0\exp[i\mathcal{S}(\phi)]$ in terms of its action $\mathcal{S}(\phi)$. Then,
 the four-momentum of such an excitation is given by
 \begin{eqnarray}\label{kas}
 k_a=\frac{1}{i\phi}(\nabla_a\mp iqA_a)\phi,
\end{eqnarray}
where $\mp q$ is the electric charge of the positive$/$negative energy excitation, respectively,  and $A_a=(-Q/r,0,0,0)$ is the 4-potential of the electromagnetic field.
Then, within the  WKB approximation let us consider  the ansatz
 \begin{eqnarray}\label{eikonal}
 \mathcal{S}(\phi)=\mp\omega v+\int^rdr'k_r(r'),
\end{eqnarray}
for the phase of the field configuration, where the top and bottom sign $\mp$ refer, respectively,  to positive and negative energy excitations.
Plugging it into (\ref{kas}), the wave four-vector takes the form,
\bqn
\lb{kv}
k_adx^a&=&\mp(\omega-q\varphi) dv+k_rdr\nb\\
&=& \big[\pm(\omega-q\varphi)\ell_{-a}+k_r\rho_a\big]dx^a,
\eqn
where $\varphi=Q/r$ is the electric potential, $\ell_{-a}=(-1,0)$ is the radial null vector, and $\rho_a=(0,1)$ is the redshift vector.
The radial momentum $k_r$ can be solved from the dispersion relation
\bq
\lb{dr}
e(r)k_r^2 \mp2(\omega-q\varphi)k_r = k^2,
\eq
once  $k^2(\omega)$  is given. Clearly, in general the above equation has four solutions: $k_{r(I)}^\pm$ and $k_{r(O)}^\pm$, where $\pm$ refer, respectively, to the positive and negative energy,
I (O) means in-going (out-going) particles. Due to the time reversal invariance, we have $k_{r(O)}^+=-k_{r(I)}^-$ and $k_{r(O)}^-=-k_{r(I)}^+$. From the standard results in quantum
mechanics, the emission rate $\Gamma$ is given by $\Gamma\sim\exp[-2$Im$\mathcal{S}]$. From Eq.(\ref{eikonal}) we can see that only the singular parts of $k_r(r)$ have
contributions to  Im$\mathcal{S}$. In particular, we have
\bqn
\lb{ImSa}
{\mbox{Im}} \mathcal{S} &=&{\mbox{Im}} \lim_{\epsilon \rightarrow 0}{\int^{r_H + \epsilon}_{r_H - \epsilon}{k_{r(O)}^+(r') dr'}}  \nb\\
&=& - {\mbox{Im}} \lim_{\epsilon \rightarrow 0}{\int_{r_H + \epsilon}^{r_H - \epsilon}{k_{r(I)}^-(r') dr'}}\nb\\
&=&  {\mbox{Im}} \lim_{\epsilon \rightarrow 0}{\int_{r_H + \epsilon}^{r_H - \epsilon}{k_{r(O)}^+(r') dr'}},
\eqn
where $r_H$ is the location of the singularity of $k_{r(O)}^+(r)$. Deforming the contour into the low half complex plane of the singularity located at $r = r_H$ for the first integral and the upper half complex plane  for the last one,
we find
\bqn
\lb{impart}
2{\mbox{Im}} \mathcal{S} &=&{\mbox{Im}} \lim_{\epsilon \rightarrow 0}\left\{{\int^{r_H + \epsilon}_{r_H - \epsilon}{k_{r(O)}^+(r') dr'}}     \right.\nb\\
&& \left. ~~~~~~~~~~~~ + {\int_{r_H + \epsilon}^{r_H - \epsilon}{k_{r(O)}^+(r') dr'}}\right\}\nb\\
&=& \text{Im}\oint drk^+_{r(O)}(r),
\eqn
where the closed circuit is always anticlockwise.  Therefore,  to calculate the emission rate we need  only consider the out-going positive energy particles.

On the other hand, in the  frame comoving with the aether, $k_a$ can be written as
\bq
\lb{kr}
k_a=-k_uu_a+k_ss_a,
\eq
where $k_u\equiv (u\cdot k)$ and $k_s\equiv (s\cdot k)$ are corresponding to, respectively,  the energy and momentum, measured by observers that are comoving with the aether, and are given by
\bqn
\label{kuks}
k_u(r)&=&\frac{\pm(\omega-q\varphi)}{(u\cdot\chi)-(s\cdot\chi)}-k_r(s\cdot\chi),\nb\\
k_s(r) &=& \frac{\pm(\omega-q\varphi)}{(u\cdot\chi)-(s\cdot\chi)}-k_r(r)(u\cdot\chi).
\eqn
Then, we have
$k^2 = - k_u^2 + k_s^2$, which is a function of $k_r$.
In this paper, we consider the non-relativistic dispersion relation, given by  \cite{unruh,WWM},
 \begin{eqnarray}\label{ksquare}
k_u^2=k_0^2\sum_{n=1}^{z}
a_n\Big(\frac{k_s}{k_0}\Big)^{2n},
\end{eqnarray}
where  $a_n$'s are dimensionless constants, which will be considered as order of unit in the following discussions \cite{WWM}, and $z$ is an integer \footnote{ A more general expression for the nonlinear dispersion relation in a curved
background was given in \cite{JB13}. However, to make the problem attackable, in this paper we restrict ourselves to the cases defined by
Eq.(\ref{ksquare}). For a further justification of the use of this form at the universal horiozn, see \cite{JB13}.}. Lorentz symmetry requires
$(a_1, z)= (1, 1)$. Therefore, in this paper we shall set $a_1 = 1$.  In the Horava theory of gravity \cite{Horava},
the power-counting renormalizability requires $z \ge 3$. The constant $k_0$ is the UV Lorentz-violating (LV) energy scale for the matter \cite{cropp} or the suppression mass scale \cite{WWM}.
The experimental viable range for the $k_0$ is rather broad and its value shows the size of LV of the given field. When $k_s/k_0\rightarrow0$, the field becomes relativistic  and one
recovers the standard dispersion relation $k_u^2 = k_s^2$.

To study the effects of high-order corrections, characterized by the critical exponent $z$, in the following we shall study the Hawking radiation for various choices of $z$ at both of the universal and Killing horizons.

 To see clearly the difference between relativistic and non-relativistic  particles, in the following we first consider the relativistic case ($z=1$), and re-obtain the well-known results of the Hawking radiation at the Killing horizons \cite{Carlip,unruh,MP09}.
However, we find that at  universal horizons relativistic particles are not created. Then, we move onto the non-relativistic ones ($z \ge 2$), and show that such particles are indeed created at universal horizons.
It should be noted that  in doing so we implicitly assume that both of these two kinds of horizons have an associated temperature. However, this is not well grounded \cite{Visser14}, and is closely related to the theory of Hawking radiation at
high energies. We shall come back to this issue at the end of Section V. In addition, in high frequencies non-relativistic particles ($z \ge 2$) are not created at Killing horizons, which confirms the earlier  findings \cite{unruh,MP09}.

\subsection{Hawking radiation for  $z=1$}

When $z=1$ or $k_s\ll k_0$, the  dispersion relation reduces to the relativistic one, $k^2=-k_u^2+k_s^2=0$, or $k_u=\pm k_s$.
From Eq.(\ref{kuks}), one can see that at both of the   Killing and universal horizons,   the solution $k_u=k_s$ will all lead to $k_r=0$.
 For the outgoing positive energy or ingoing negative energy particles, the relation $k_u=-k_s$ together with Eq.(\ref{kuks})  leads to
\begin{eqnarray}\label{krkilling}
 k_{r(O)}^+(r)&=&-\frac{2(\omega-q\varphi)}{(s\cdot\chi)
-(u\cdot\chi)}\frac{1}{(s\cdot\chi)
+(u\cdot\chi)}\nb\\
&=& \frac{2(\omega-q\varphi)}{e(r)},
\end{eqnarray}
which is finite  at the universal horizon $(u\cdot\chi)=0$, but singular  at the Killing horizon $e(r) = 0$. This  implies  that {{\em relativistic particles   cannot escape from the universal horizons even quantum mechanically, as their velocity is finite  and
the horizon serves as an infinitely large barrier to them}.   However,  they
can be  created at the Killing horizon with the standard results \cite{Carlip}, }
\bqn
\label{imkh}
&&  2\text{Im} S =\frac{\omega-\mu_0}{T_{KH}},\nb\\
 \lb{tkh}
&& T_{KH}=\frac{e'(r_{KH})}{4\pi} = \frac{\kappa_{KH}^{GR}}{2\pi}, 
\eqn
where $ \mu_0=q\varphi_{KH} $ and $ \varphi_{KH} \equiv Q/r_{KH}$, a prime  denotes the derivative with respect  to $r$, and
$\kappa_{KH}^{GR}$ denotes the surface gravity defined  as
\bq
\lb{surfaceGKH}
 \kappa^{GR} \equiv \sqrt{-\frac{1}{2}\left(\nabla_{a}\chi_{b}\right)\left(\nabla^{a}\chi^{b}\right)}\;.
\eq

It should be noted that, in Ref. \cite{michel} by using collapsing shell method, the authors showed that at the Killing horizon, with a given $k_0$  there exists an effective
temperature $T_\omega(k_0)$. When $k_0$ is  increasing, $T_\omega$ approaches to the   Hawking temperature $T_{KH}$. In Ref. \cite{cropp}, on the other hand,  it
was shown that energetic particles  simply pass  the Killing horizon, while low-energy  particles linger and eventually  escape to infinity.

\subsection{Hawking radiation for $z> 1$}

When $z>1$, from Eq.(\ref{kuks}) we find that,
\begin{eqnarray}\label{kukr}
&&k_u(r)=\frac{1}{(u\cdot\chi)}\big[\pm(\omega-q\varphi)
+k_s(r)(s\cdot\chi)\big],\nb\\
&&
k_r(r)=-\frac{1}{(u\cdot\chi)}\left[\frac{\mp(\omega-q\varphi)}
{(u\cdot\chi)-(s\cdot\chi)}
+k_s(r)\right].  ~~~~~
\end{eqnarray}
At the Killing horizon we have $(s\cdot\chi)=-(u\cdot\chi)$,  and $(u\cdot\chi)$ is finite,
so one can see that the momentum $k_r$ is
always regular, indicating that non-relativistic particles may not created at the Killing horizon, as they can escape the Killing horizon even classically.
This is consistent with the results obtained in \cite{unruh,MP09}. The reason is simply the following: To have terms with $z > 1$ be leading, we implicitly assume that
$k > k_0$, as one can see from Eq.(\ref{ksquare}). Therefore, our above claim is actually valid only for modes with $k > k_0$, i.e., the high frequency modes \cite{unruh,MP09}.
For modes with $k < k_0$, the quadratic term $k^2$ is important, and we must consider it together with high-order corrections. In the latter, it was shown that the spectrum of the corresponding
Hawking radiation is modified \cite{unruh,MP09}. So,  in the rest of this section we shall focus ourselves  only at the universal horizon.

For the outgoing modes with  positive Killing energy  [the top sign in Eqs.(\ref{kukr})], $k_s(r)$ has a singularity at the universal horizon. In review of Eqs.(\ref{dr}),
(\ref{kuks}) and (\ref{ksquare}),  we assume that it takes the    form
\begin{eqnarray}
\label{ks}
k_s(r)=\frac{k_0b(\omega,r)}{\left|u\cdot\chi\right|^m
},\quad m>0,
\end{eqnarray}
where $b\left(\omega,r_{UH}\right)\neq 0$, and $m$ is the smallest positive real number such that $\left|u\cdot\chi\right|^m k_s(r)$ is finite at the horizon.
Combining  Eq.(\ref{ks}) with Eqs.(\ref{ksquare}) and (\ref{kukr}), we find that $m=1/(z-1)$. Then, the outgoing positive energy mode is given by,
 \begin{eqnarray}
 \label{krp}
k_{r(O)}^+(r)&=&\frac{1}{(-u\cdot\chi)} \left[\frac{\omega-q\varphi}{
(s\cdot\chi-u\cdot\chi)}
+\frac{k_0b}{\left|u\cdot\chi\right|^{\frac{1}{z-1}}}\right],\nb\\
\end{eqnarray}
where $b$ satisfies the relation
 \begin{eqnarray}\label{brelation}
b\left[\sqrt{a_z}b^{z-1}-(s\cdot\chi)\right]=\frac{\omega-q\varphi}{k_0}{\left|u\cdot\chi\right|^{\frac{1}{z-1}}}.
\end{eqnarray}

In the following, let us consider the three cases,  $z=2$, $z=3$ and $z\ge 4$, separately.

\subsubsection{Hawking radiation with $z=2$}

{ This case was studied in some detail in \cite{JB13}, and results for $Q = 0$ were  reported in \cite{berglund2}.
To show how to generalize such studies to the cases with $z > 2$, in the following let us first study  this case in more details.}
 In particular, when $z =2$, we have   $m=1/(z-1) =1$. It can be shown that this is  the only case in which  $m$ is an integer.
 Then,
Eqs.(\ref{krp}) and (\ref{brelation}) become
\begin{eqnarray}\label{krb2a}
&&k_{r(O)}^+(r)=\frac{\omega-q\varphi}{(-u\cdot\chi)
(s\cdot\chi-u\cdot\chi)
}+\frac{k_0b}{(- u\cdot\chi)^2
}, ~~~~~~\\
&&
\label{krb2}
b\left[\sqrt{a_{2}}b-(s\cdot\chi)\right]=\frac{\omega-q\varphi}{k_0}(-u\cdot\chi).
\end{eqnarray}
Denoting $\epsilon \equiv r-r_{UH}$, we find that  near the universal horizon $r=r_{UH}$ we have
\begin{eqnarray}\label{us2}
&&(-u\cdot\chi)=\epsilon \left[\alpha_1+\alpha_2\epsilon +\mathcal{O}(\epsilon ^2)\right],\nonumber\\
&&
(s\cdot\chi)=s_0+s_1\epsilon +\mathcal{O}(\epsilon ^2),
\eqn
where
\bqn
&& \alpha_1\equiv(-u\cdot\chi)'|_{{UH}}>0,\quad
\alpha_2\equiv\frac{1}{2}(-u\cdot\chi)''|_{{UH}}<0,
 \nonumber\\&&s_0\equiv(s\cdot\chi)|_{{UH}},\quad s_1\equiv(s\cdot\chi)'|_{{UH}}.
\end{eqnarray}
Setting
\begin{eqnarray}\label{udx2}
&&b=b_0+b_1\epsilon +\mathcal{O}(\epsilon ^2),
\end{eqnarray}
from Eq.(\ref{krb2}), we obtain
\begin{eqnarray}\label{b1g}
&& b_0=\frac{s_0}{\sqrt{a_2}}, \quad b_1=\frac{\omega-q\varphi}{s_0k_0}\alpha_1+\frac{s_1}{\sqrt{a_2}}.
\end{eqnarray}

On the other hand, we also have,
\begin{eqnarray}\label{usx2}
(-u\cdot\chi)^{-2}&=&\frac{1}{\epsilon ^2}
\left(\frac{1}{\alpha_1+\alpha_2\epsilon +\mathcal{O}(\epsilon ^2)}\right)
^{2}\nonumber\\
&=&\frac{1}{\epsilon ^2}\left(\frac{1}{\alpha_1}-\frac{\alpha_2}{\alpha_1^2}\epsilon
+\mathcal{O}(\epsilon ^2)\right)^2\nonumber\\
&=&\frac{1}{\epsilon ^2}\left(\frac{1}{\alpha_1^2}-2\frac{\alpha_2}{\alpha_1^3}\epsilon
+\mathcal{O}(\epsilon ^2)\right).
\end{eqnarray}
Substituting it together with Eq.(\ref{udx2}) into  Eq.(\ref{krb2a}), we find,
\begin{eqnarray}\label{krexpand2}
&& k_{r(O)}^+(r)\simeq2\frac{\omega-q\varphi-\mu}{s_0\alpha_1
}\frac{1}{\epsilon }+\frac{k_0b_0}{(\alpha_1\epsilon )^2},\nb\\
&& \mu= -\frac{k_0}{2}\left(\frac{s\cdot\chi}{a\cdot s}\right)\left[\frac{(s\cdot\chi)'}{\sqrt{a_{2}}}
+\frac{(s\cdot\chi)(u\cdot\chi)''}{\sqrt{a_{2}}(a\cdot s)}\right]_{UH}. ~~~~~~~~~
\end{eqnarray}
Inserting the above expressions into Eq.(\ref{impart}), and using the residual theorem, we finally obtain the Boltzman factor
\begin{eqnarray}
 &&2\text{Im} S =\frac{\omega-\mu_0}{T^{z=2}_{UH}},
\end{eqnarray}
where $\mu_0=(q\varphi+\mu)_{UH}$ is the chemical potential of the scalar field, and
\begin{eqnarray}\label{temperature2}
 && \left. T^{z=2}_{UH}=\frac{(a\cdot s)(s\cdot\chi)}{4\pi }\right|_{UH} 
 =  \frac{\kappa_{UH}}{2\pi},
\end{eqnarray}
where $\kappa_{UH} = s_0\alpha_1/2$ denotes the surface gravity defined by Eq.(\ref{surfaceG}). Clearly, $T^{z=2}_{UH}$ and $\kappa_{UH}$ satisfy the standard relation
$T = \kappa/2\pi$ \cite{cropp}. However, as to be shown below, this is no longer  the case for a general $z$, although $T^{z}_{UH}$ is still proportional to
$\kappa_{UH}$.

Applying the above general formula (\ref{temperature2}) to the two particular solutions given in the last section, we find that
\bqn
\lb{tuh}
T^{z=2}_{UH} = \begin{cases}
 \frac{1}{4\pi r_{UH}\sqrt{3\mathrm{g}}}\sqrt{(1-\frac{Q^2}{r_{UH}^2})
 (2-\frac{Q^2}{r_{UH}^2})}, & c_{14} = 0,\cr
 \frac{1}{4\sqrt{\mathrm{g}}\pi r_{UH}}\sqrt{\mathrm{p}-
 \frac{Q^2}{r^2_{UH}}}, & c_{123} = 0.  \cr
 \end{cases}\nb\\
\eqn
 When $Q=0$,  it reduces to the one obtained in \cite{berglund2}, calculated in the PG coordinates. 
 However, it is interesting to note
that such obtained temperature is different from that obtained by the Smarr relation, by simply adopting the mass defined
in \cite{eling}. We shall come back to this issue in the next section.

\subsubsection{Hawking radiation with $z=3$}

In the Horava theory \cite{Horava}, the power-counting renormalizability condition  requires $z\ge3$, as mentioned above. Therefore, the case $z = 3$ has particular interest, as far as the Horava theory
is concerned.

When $z \ge 3$ the parameter $m [\equiv 1/(z-1)]$ introduced in Eq.(\ref{ks}) can no longer be an integer, and the nature of the singularity at $u\cdot\chi = 0$ becomes a  branch point, instead of a single pole.
 To handle this case carefully,  we shall use two different methods. One is the more ``traditional" one, and  the other is the so-called fractional derivative, a  branch of mathematics, which
has  already been well-established  \cite{uchaikin} and  applied to physics in similar situations  in various occasions  \cite{calcagni}. We shall show that both methods yield the same results,
 as it should be expected.

Let us first consider  the quantity $\left|u\cdot\chi\right|^{m}$, for which we find that it is easier to consider the regions  $r>r_{UH}$ and $r<r_{UH}$, separately. In particular,   in the region $r>r_{UH}$ we have $(u\cdot\chi)<0$.
Then, Eqs.(\ref{krp}) and (\ref{brelation}) become
\begin{eqnarray}\label{krb3}
&&k_{r(O)}^+(r)=\frac{\omega-q\varphi}{(-u\cdot\chi)
(s\cdot\chi-u\cdot\chi)
}+\frac{k_0b}{(- u\cdot\chi)^{3/2}
}, ~~~~~~~~~\\
\label{krb3a}
&&
b\left[\sqrt{a_{3}}b^2-(s\cdot\chi)\right]=\frac{\omega-q\varphi}{k_0}\left({-u\cdot\chi}\right)^{1/2}.
\end{eqnarray}

At the universal horizon, we have $\left({-u\cdot\chi}\right) \propto \epsilon $ to the leading order of $\epsilon $.  Then, the leading term of the right-hand side of
Eq.(\ref{krb3a}) is  proportional to $\epsilon ^{1/2}$. This implies that the function $b(r)$ must be expanded in terms of  $\epsilon ^{1/2}$, instead of $\epsilon $ as done in the last
case with $z = 2$. So, setting
\begin{eqnarray}
\label{bexpand}
&&b=b_0+b_1\epsilon ^{1/2}+b_2\epsilon +b_3\epsilon ^{3/2}\nb\\
&&~~~~~+b_4\epsilon ^{2}+b_5\epsilon ^{5/2}+\mathcal{O}(\epsilon ^3),
\end{eqnarray}
we can determine the coefficients $b_i$'s from the relation,
\begin{eqnarray}
&&b^2\left[\sqrt{a_{3}}b^{2}-(s\cdot\chi)\right]^2 = \frac{(\omega-q\varphi)^2}{k_0^2}\left({-u\cdot\chi}\right),  ~~~~
\end{eqnarray}
which  yields,
 \begin{eqnarray}\label{b13}
 &&b_0=\left(\frac{s_0}{\sqrt{a_3}}\right)^{1/2},\quad b_1=\frac{\sqrt{\alpha_1}(\omega-q\varphi)}{2s_0k_0},\nb\\
 &&b_2=\frac{4{s_0}^{2}{k_0}^{2}s_1-3\alpha_1\sqrt{a_3}(\omega-q\varphi)^2}{8~{a_3}^{\frac{1}{4}}{k_0}^{2}{s_0}^{\frac{5}{2}}},\nb\\
 &&b_3=\frac{\omega-q\varphi}{{4\sqrt{\alpha_1}{k_0}^{3}{s_0}^{4}}}[{k_0}^{2}{s_0}^{2}(\alpha_2s_0-2s_1\alpha_1)\nb\\
 &&~~~~~~~~~~~~~~~~~~~~~~~+2\sqrt{a_3}\alpha_1^2(\omega-q\varphi)^2].\nb\\
 \end{eqnarray}
 From the above derivation,  it is easy to see that, if the term $b_1\epsilon ^{1/2}$ were not present,
 Eq.(\ref{krb3a}) would not hold.

 To calculate the last term appearing in the right-hand side of Eq.(\ref{krb3}), as mentioned above, we use two different methods. Let us first consider the fractional derivative. Since $\lim_{\epsilon  \rightarrow 0}
\int{\epsilon ^{\delta} d\epsilon  } = 0$ for any $\delta > -1$, we need to consider the fractional expansion of Eq.(\ref{A.1}) only up to $\epsilon ^{-3/2}$, which is sufficient for the calculation of
$2{\mbox{Im}} \mathcal{S}$ given by Eq.(\ref{impart}). Then, from Eq.(\ref{us2}) and
Eqs.(\ref{A.1}) - (\ref{A.3})  we find that, after taking $\alpha = 1/(z-1) = 1/2$, $\left({-u\cdot\chi}\right)^{- 3/2}$ is given by
\bq
\lb{u3}
\left({-u\cdot\chi}\right)^{- 3/2} = \epsilon ^{-3/2}\left(\alpha_1^{-3/2} +  \mathcal{O}(\epsilon )\right).
\eq
This can be also obtained from the following considerations. First,  from Eq.(\ref{krb3a}) we have
\begin{eqnarray}
\left({-u\cdot\chi}\right)^{3/2} =\left(\frac{k_0}{\omega-q\varphi}\right)^3 b^3\left[\sqrt{a_{3}}b^2-(s\cdot\chi)\right]^{3}. ~~~~
\end{eqnarray}
Substituting   Eqs.(\ref{bexpand})-(\ref{b13}) into the right-hand side of the above expression,  we obtain
\begin{eqnarray}
({-u\cdot\chi})^{3/2}=\epsilon ^{3/2}\left(\alpha_1^{3/2} +  \mathcal{O}(\epsilon )\right).
\end{eqnarray}
 Assuming that $({-u\cdot\chi})^{-3/2}$ takes the form, $({-u\cdot\chi})^{-3/2}=\hat{a}_1\epsilon ^{-3/2}  +  \mathcal{O}\left(\epsilon ^{-1/2}\right)$, then, using the identity $({-u\cdot\chi})^{3/2} \cdot ({-u\cdot\chi})^{-3/2}  = 1$, we find that
$({-u\cdot\chi})^{-3/2}$ is precisely given by Eq.(\ref{u3}).

Substituting Eqs.(\ref{bexpand}) and (\ref{u3}) into  Eq.(\ref{krb3}), we find,
\begin{eqnarray}\label{krexpand3}
k_{r(O)}^+&=&\frac{\omega-q\varphi}{s_0
}\frac{1}{\epsilon [\alpha_1+\mathcal{O}(\epsilon )]}
+\frac{k_0[b_0+b_1\epsilon ^{1/2}+\mathcal{O}(\epsilon )]
}{\epsilon ^{3/2}[\alpha_1+\mathcal{O}(\epsilon )]^{3/2}
}\nb\\
&\simeq&\frac{3}{2}\frac{\omega-q\varphi}{s_0\alpha_1
}\frac{1}{\epsilon }+\frac{k_0b_0}{(\epsilon \alpha_1)^{3/2}}.
\end{eqnarray}

In the  region $r<r_{UH}$ we have $(u\cdot\chi)>0$, and Eqs.(\ref{krp}) and (\ref{brelation}) become
\begin{eqnarray}
\label{krexpand3s}
k_{r(O)}^+&=&\frac{\omega-q\varphi}{s_0
}\frac{1}{-\epsilon[\alpha_1+\mathcal{O}(\epsilon)]}
-\frac{k_0[b_0+b_1\epsilon^{1/2}+\mathcal{O}(\epsilon)]
}{\epsilon^{3/2}[\alpha_1+\mathcal{O}(\epsilon)]^{3/2}
}\nb\\
&\simeq&\frac{3}{2}\frac{\omega-q\varphi}{s_0\alpha_1
}\left(-\frac{1}{\epsilon}\right)-\frac{k_0b_0}{(\epsilon\alpha_1)^{3/2}}.
\end{eqnarray}

We set $\epsilon\equiv r_{UH}-r$ and
following a similar  procedure, it can be shown that
\begin{eqnarray}\label{krexpand3s}
k_{r(O)}^+&=&\frac{\omega-q\varphi}{s_0
}\frac{1}{-\epsilon[\alpha_1+\mathcal{O}(\epsilon)]}
-\frac{k_0[b_0+b_1\epsilon^{1/2}+\mathcal{O}(\epsilon)]
}{\epsilon^{3/2}[\alpha_1+\mathcal{O}(\epsilon)]^{3/2}
}\nb\\
&\simeq&\frac{3}{2}\frac{\omega-q\varphi}{s_0\alpha_1
}\left(-\frac{1}{\epsilon}\right)-\frac{k_0b_0}{(\epsilon\alpha_1)^{3/2}}.
\end{eqnarray}
Setting $r=r_{UH}+\epsilon  e^{i\theta}$, we find
\begin{eqnarray}\label{}
k_{r(O)}^+
&\simeq&\frac{3}{2}\frac{\omega-q\varphi}{s_0\alpha_1
}\frac{1}{\epsilon  e^{i\theta}}+\frac{k_0b_0}{(\epsilon  e^{i\theta}\alpha_1)^{3/2}}.
\end{eqnarray}
Inserting the above expression into Eq.(\ref{impart}),
we find
\begin{eqnarray}
 2\text{Im}\mathcal{S}= \frac{\omega - q\varphi -\mu}{T^{z=3}_{UH}},
 \end{eqnarray}
 where
 \begin{eqnarray}
  T^{z=3}_{UH}&=&\left. \frac{(a\cdot s)(s\cdot\chi)}{3\pi }\right|_{UH} = \frac{2\kappa_{UH}}{3\pi},\nb\\
\mu &\equiv& - T^{z=3}_{UH} {\cal{I}},
\end{eqnarray}
with $dr = i\epsilon  e^{i\theta} d\theta$,  and
\bq
\lb{I2}
{\cal{I}} \equiv  \text{Im}\lim_{\epsilon \rightarrow 0} \oint dr\frac{k_0b_0}{(\epsilon  e^{i\theta}\alpha_1)^{3/2}}.
\eq
To calculate  ${\cal{I}} $, we first note that
\bq
\lb{complexA}
\left(e^{i\theta}\right)^n = e^{i n \theta},\;\;\;\;
 \left(e^{i\theta}\right)^{1/n} = e^{i(\theta + 2m\pi)/n},
 \eq
 where $n$ is an integer, and $m = 0, 1, 2, ..., n-1$. Then, we find that
 \begin{eqnarray}
{\cal{I}} &=&\text{Im}\lim_{\epsilon  \rightarrow0}\int_0^{2\pi} \frac{ik_0b_0 \epsilon  e^{i\theta}}{(\epsilon  e^{i\theta}\alpha_1)^{3/2}}d\theta\nb\\
 &=&\text{Im}\lim_{\epsilon  \rightarrow0}\left(\frac{ik_0b_0}{\sqrt{\epsilon }\alpha_1^{3/2}}\int_0^{2\pi}  e^{-i(\theta + 6m\pi)/2}d\theta\right)\nb\\
&=&\text{Im}\lim_{\epsilon  \rightarrow0}\left((-1)^{m}\frac{4k_0b_0}{\sqrt{\epsilon }\alpha_1^{3/2}}\right) =0.
\end{eqnarray}
Thus, finally we  obtain
\begin{eqnarray}
 2\text{Im}\mathcal{S}=\frac{\omega-q\varphi}{T^{z=3}_{UH}}.
 \end{eqnarray}

It is interesting to note that $T^{z=3}_{UH}$ given above is larger than $T^{z=2}_{UH}$ by a factor
$4/3$, although both of them are proportional to the surface gravity $\kappa_{UH}$ defined by Eq.(\ref{surfaceG}). In addition, the real part of ${\cal{I}} $ diverges, although
its imaginary part vanishes. This is similar to the extremal black holes \cite{KM}, which are considered to be able  in thermal equilibrium at any finite  temperature \cite{HHR}.

\subsubsection{Hawking radiation   with   $z \ge 4$}

With the above preparations, we are ready to consider the general case with any given $ z\ge 4$.   Similar to the case $z = 3$, let us first consider the region $r>r_{UH}$,
in which we have $(u\cdot\chi)<0$, and Eqs.(\ref{krp}) and (\ref{brelation}) become
\begin{eqnarray}\label{krb2n}
&&k_{r(O)}^+(r)=\frac{\omega-q\varphi}{(-u\cdot\chi)
(s\cdot\chi-u\cdot\chi)
}+\frac{k_0b}{(- u\cdot\chi)^{\frac{z}{z-1}}
},\nonumber\\&&
b\left[\sqrt{a_{z}}b^{z-1}-(s\cdot\chi)\right]=\frac{\omega-q\varphi}{k_0}(-u\cdot\chi)^{\frac{1}{z-1}}.
\end{eqnarray}

To obtain the function $b(\omega, r)$, we need to expand $(-u\cdot\chi)$ only  to the first order of $\epsilon $. So, from Eq.(\ref{krb2n}) we find
\begin{eqnarray}\label{udx}
(-u\cdot\chi)^{\frac{1}{z-1}}&=&\left[\alpha_1\epsilon +
\mathcal{O}\left(\epsilon ^2\right)\right]^{\frac{1}{z-1}}\nonumber\\
&=&(\alpha_1\epsilon )^{\frac{1}{z-1}}+
\mathcal{O}\left(\epsilon ^{\frac{2}{z-1}}\right).
\end{eqnarray}
Therefore, For any given  $z$, the following expansion must be performed,
\begin{eqnarray}\label{f}
&&b=b_0+b_1\epsilon ^{\frac{1}{z-1}}+\mathcal{O}\left(\epsilon ^{\frac{2}{z-1}}\right).
\eqn
Substituting Eqs.(\ref{udx}) and (\ref{f}) into Eq.(\ref{krb2n}), we get
\begin{eqnarray}\label{b1}
&&b_0=\left(\frac{s_0}{\sqrt{a_{z}}}\right)^{\frac{1}{z-1}},\quad b_1= {\frac{1}{z-1}}\frac{\omega-q\varphi}{s_0k_0}
\alpha_1^{\frac{1}{z-1}}. ~~~~~~~~~~
\end{eqnarray}
Hence, we obtain
\begin{eqnarray}\label{krexpand2n}
&&k_{r(O)}^+(r)\simeq \left(\frac{z}{z-1}\right)\frac{\omega-q\varphi}{s_0\alpha_1
}\frac{1}{\epsilon }+\frac{k_0b_0}{(\epsilon \alpha_1)^{\frac{z}{z-1}}}.  ~~~~~~~~
\end{eqnarray}
It is interesting to note the $z$-dependence of $k_{r(O)}^+(r)$. In addition, as in the last case, the above expression for $k_{r(O)}^+(r)$ can be obtained by either the fractional derivative with
$\alpha = 1/(z-1)$ or the more traditional method, illustrated above.

In the region $r<r_{UH}$, we have $(u\cdot\chi)>0$, and Eqs.(\ref{krp}) and (\ref{brelation}) become
\begin{eqnarray}\label{krbodd}
&&k_{r(O)}^+(r)=\frac{\omega-q\varphi}{(-u\cdot\chi)
(s\cdot\chi-u\cdot\chi)
}-\frac{k_0b}{( u\cdot\chi)^{\frac{z}{z-1}}
},\nonumber\\&&
b\left[\sqrt{a_{z}}b^{z-1}-(s\cdot\chi)\right]=\frac{\omega-q\varphi}{k_0}(u\cdot\chi)
^{\frac{1}{z-1}}.
\end{eqnarray}

Following the same steps as given in the region $r>r_{UH}$ we find that,
\begin{eqnarray}\label{krexpandodd}
&&k_{r(O)}^+(r)\simeq \left(\frac{z}{z-1}\right)\frac{\omega-q\varphi}{s_0\alpha_1
}\left(-\frac{1}{\epsilon}\right)-\frac{k_0b_0}
{(\epsilon\alpha_1)^{\frac{z}{z-1}}}.\nb\\
\end{eqnarray}
Combining Eqs.(\ref{krexpand2n}) and (\ref{krexpandodd}),  and let $r=r_{UH}+\epsilon  e^{i\theta}$, there has
\begin{eqnarray}\label{}
k_{r(O)}^+
&\simeq&\frac{z}{z-1}\frac{\omega-q\varphi}{s_0\alpha_1
}\frac{1}{\epsilon  e^{i\theta}}+\frac{k_0b_0}{(\epsilon  e^{i\theta}\alpha_1)^{\frac{z}{z-1}}}.
\end{eqnarray}
Considering Eq.(\ref{impart}),
we find that
\bq
\lb{Tz}
 2\text{Im}\mathcal{S}= \frac{\omega - q\varphi -\mu}{T^{z\ge 4}_{UH}},
 \eq
 where
 \bqn
  T^{z\ge 4}_{UH}  &=& \frac{(z-1)s_0\alpha_1}{2\pi z} = \frac{2(z-1)}{z}  T^{z = 2}_{UH},\nb\\
  \mu &=& - T^{z\ge 4}_{UH}\; {\cal{I}}_z,
  \eqn
  with
 \begin{eqnarray}
{\cal{I}}_z &\equiv& \text{Im}\lim_{\epsilon \rightarrow0}\int_0^{2\pi} \frac{ik_0b_0\epsilon  e^{i\theta}}{(\epsilon  e^{i\theta}\alpha_1)^{\frac{z}{z-1}}}d\theta\nb\\
&=&\text{Im}\lim_{\epsilon \rightarrow0}\left(\frac{ik_0b_0}{\left(\alpha^z_1\epsilon \right)^{\frac{1}{z-1}}}
\int_0^{2\pi}e^{-i(\theta+2zm\pi)/(z-1)}d\theta\right)\nb\\
&=& \text{Im}\lim_{\epsilon \rightarrow0}\left[\frac{(1-z)k_0b_0}{
\left(\alpha^z_1\epsilon \right)^{\frac{1}{z-1}}}
e^{-i2\pi \frac{mz}{z-1}}\left(e^{-\frac{i2\pi}{z-1}}-1\right)\right]\nb\\
&=& \lim_{\epsilon \rightarrow0}\left\{\frac{2(z-1)k_0b_0}{
\left(\alpha^z_1\epsilon \right)^{\frac{1}{z-1}}}\sin\frac{\pi}{z-1}\cos\frac{(2m+1)\pi}{z-1}\right\}\nb\\
&=& \begin{cases} 0, & z= \infty,\cr
\pm\infty, & 4\leq z< \infty, \cr
\end{cases}
\end{eqnarray}
where $m = 0, 1,  ..., z-2$, and
\bq
\lb{sing}
\pm = \mbox{Sign}\left\{\cos\left(\frac{(2m+1)\pi}{z-1}\right)\right\}.
\eq
 Thus,  the chemical potential for $4\leq z< \infty$ is always unbounded, unless $z = \infty$. In the latter, similar to the cases $z = 2$ and $z= 3$, it vanishes.
 It is interesting to note that the signs of ${\cal{I}}_z$ depends  not only on $z$ but also on $m$. In particular, when $m = 0$ and $m = z-2$, $ \cos[(2m+1)\pi/(z-1)]$ is always positive, so that $\mu \propto - {\cal{I}}_z$ always approaches to $-\infty$.
 Therefore, for any given $z$ there always  exists an intermediate region    in which $\mu$ always approaches to $+ \infty$. One may consider this range   as
 physically not realizable, as the corresponding chemical potential becomes infinitely large.

As noted previously, the temperature of the universal horizon is always finite and depends on $z$ explicitly, which characterizes another feature of the nonlinear dispersion relation.
Therefore, although, to the leading order, the Hawking radiation is thermal for any given species with a fixed $z$,
the temperature of such a species depends explicitly on $z$, and  increases as $z$ increases. In particular, as $z \rightarrow \infty$, a particular case considered
also in  \cite{JB13}, it approaches to its maximum $T^{z=\infty}_{UH} = 2T^{z=2}_{UH}$.

\section{Modified Smarr formula and Mass of a Black Hole}
\renewcommand{\theequation}{4.\arabic{equation}} \setcounter{equation}{0}

{ From the above sections one can see that the Hawking radiation of non-relativistic particles can occur at the universal horizon. Then, a natural question is wether the first law of black hole mechanics also holds there?
In the neutral case, Berglund {\it et al} \cite{berglund2} found that a slightly modified first law indeed exists. But,  recently  Ding {\it et al}  $\;$ found that a simple generalization of such a formula   to the  charged
case is not possible \cite{ding}. A fundamental question is how to define the entropy at the universal horizon, although it is quite reasonable to assume that such an entropy exists.  Indeed,
from Wald's entropy formula \cite{wald1994},  it was shown that  the  entropy $S$ of the universal horizon is still proportional to its area $S=A_{UH}/4$ \cite{Mohd}, since none of the  terms $\mathcal{L}_{\ae}$
and $\mathcal{L}_{M}$ appearing  in Eq.(\ref{lag})  depends on the curvature $R_{\mu\nu\alpha\beta}$.

In this section, we shall flip the logics, and assume that the entropy is proportional to the area of the universal horizon, then study the implications of the first law of black hole mechanics. In particular, we would like to find the
mass of the black hole, and then compare it with the well-known one \cite{eling,garfinkle}. The inconsistence of these two different masses imply that at least one of our assumptions needs to be modified  \footnote{It is also possible that
the masses obtained in \cite{eling,garfinkle} need to be modified.}, that is,  either the entropy is not proportional to the area of the universal horizon, or the first law of black hole mechanics at the universal horizon must be generalized, or both.

With  the temperature $T_{UH}$  of the black hole at the universal horizon calculated in the last section, and the assumption that the  entropy $S$ of the universal horizon is still proportional to its area $S=A_{UH}/4$ \cite{Mohd},}
we can uniquely determine the mass of the black hole, by assuming that   the first law of the black hole thermodynamics,
\begin{eqnarray}
\lb{1law}
 &&dM=TdS+VdQ,
\end{eqnarray}
holds at the universal horizon $r = r_{UH}$. To this purpose, let us first note that $M=M(S,Q),\;T=T(S,Q)$ and $V=V(S,Q)$, where $S = \pi r_{UH}^2$. Then, from the integrability condition
\begin{eqnarray}
\lb{InTCD}
 \frac{\partial V(S, Q)}{\partial S} =  \frac{\partial T(S, Q)}{\partial Q},
\end{eqnarray}
we find
\begin{eqnarray}\label{potential}
 V=\int\frac{\partial T(S, Q)}{\partial Q}dS + V_{o}(Q),
\end{eqnarray}
where $V_{o}(Q)$ is a function  of $Q$, and will be determined by the integrability condition (\ref{InTCD}). When $Q = 0$, we must have  $V(S, 0)=0$.
Once $V$ is known, from Eq.(\ref{1law}) we can calculate the mass of the black hole,
\begin{eqnarray}
\lb{BHmass}
 M(S, Q) =\int_0^S T(S',0)dS' + \int_0^QV(S,Q')dQ'.
\end{eqnarray}

Applying the above formulas to the two particular cases, $c_{123} = 0$ and $c_{14} = 0$, we shall obtain the mass of the black hole in each case. For the sake of simplicity,
let us consider only the case with $z = 2$.

 \subsection{Mass of the Black Hole for $c_{123} = 0$}

When $c_{123} = 0$, from Eqs.(\ref{tuh}) and (\ref{potential}) we find that
\begin{eqnarray}
\lb{potentialA}
 V=\frac{1}{2\sqrt{1-c_{13}}} \arctan\left(\frac{Q}{2\sqrt{1-c_{13}} r_{UH} {\cal{S}}}\right),
\end{eqnarray}
where
\bq
\lb{4.1}
 {\cal{S}} \equiv \sqrt{1 - \frac{c_{14}}{2} -\frac{Q^2}{r^2_{UH}}}.
 \eq
Then, Eq.(\ref{BHmass}) yields,
\begin{eqnarray}
\lb{BHmassA}
 &&M=  r_{UH} {\cal{S}} +VQ,
\end{eqnarray}
which takes precisely the Smarr form,
\bq
\lb{MassA}
M = 2T_{UH}S + VQ,
\eq
where $T_{UH}$ is given by Eq.(\ref{tuh}). It is interesting to note that the above Smarr mass
is quite different from the  {total mass}, calculated at spatial infinity \cite{ding,eling,garfinkle},
\bq
\lb{ADMmass}
M_{tot} = \left(1 - \frac{c_{14}}{2}\right)r_{UH}.
\eq

 \subsection{Mass of the Black Hole for $c_{14} = 0$}

 In this case, we find that
 \begin{eqnarray}
 V=\frac{1}{\sqrt{3(1-c_{13})}}
\left[E\left(\phi,\frac{1}{2}\right)-\frac{1}{4}F\left(\phi,\frac{1}{2}\right)\right],
\end{eqnarray}
where $\phi=\arcsin(Q/r_{UH})$, and $F$ and $E$ are,  respectively,  the first and second kind of the elliptic functions.
Then, from Eq.(\ref{BHmass}) we obtain
\begin{eqnarray}
\lb{BHmassB}
M={\cal{S}} r_{UH} +VQ,
 \eqn
 but now with
 \bqn
 {\cal{S}} \equiv  \frac{1}{\sqrt{3(1-c_{13})}} \sqrt{\left(1-\frac{Q^2}{r^2_{UH}}\right)\left(1-\frac{Q^2}{2r^2_{UH}}\right)}.
\end{eqnarray}
Again, such obtained mass satisfies the Smarr formula (\ref{MassA}).
Note that in the present case the { total mass} is given by \cite{ding,eling,garfinkle},
\bq
\lb{ADMmassB}
M_{tot} = \frac{2}{3} r_{UH} +\frac{Q^2}{3r_{UH}},
\eq
which is also different from that given by Eq.(\ref{BHmassB}).

\section{Conclusions}
\renewcommand{\theequation}{5.\arabic{equation}} \setcounter{equation}{0}

In this paper, we have studied the quantum tunneling of both relativistic and non-relativistic particles at the Killing and universal horizons of the  Einstein-Maxwell-aether black
holes found recently in \cite{ding}, by using the   Hamilton-Jacobi method  \cite{agheben,ding1,ding2}. Assuming that the dispersion relation in general takes the form
(\ref{ksquare})  \cite{unruh,WWM}, we have found that  in high frequencies only  relativistic particles ($z  =1$) can be created at the Killing horizons.
{ The radiation at the Killing horizons  is
 thermal with a temperature given by
$T^{z=1}_{KH} = {\kappa^{GR}_{KH}}/{2\pi}$} \cite{Carlip}.
This is consistent with previous results \cite{unruh,MP09}.
To the leading order, these results are also consistent with the ones obtained by studying ray trajectories \cite{cropp}, in which it was shown that $ \kappa^{GR}$ receives corrections starting from
the order of $\left(\ell \Omega\right)^{2/3}$, where $\Omega$ denotes the Killing energy at infinity, and $\ell$ is the UV Lorentz-violating scale.

On the other hand, particles with $z \ge 2$ cannot be created at Killing horizons (for high frequency modes).
If they exist right inside of a Killing horizon, they just simply pass through it and escape to infinity even classically. On the other hand, the Hawking radiation is purely quantum mechanical.
It should be noted that in \cite{cropp} it was  found that low-energy particles linger close to the Killing horizon before escaping  out to infinity,
which cannot be   seen from the current calculations of quantum tunneling.

{ At the universal horizon, the situation is different: } only non-relativistic particles (with $k > k_0$) are created quantum mechanically at the universal horizons and radiated out to infinity. The corresponding Hawking radiation is thermal, but
different species of particles, characterized by the parameter $z$,  experience different temperatures, given by
\bq
\lb{5.1}
T_{UH}^{z\ge 2}  = \left(2 - \frac{2}{z}\right)\frac{\kappa_{UH}}{2\pi},
\eq
where $\kappa_{UH}$ is the surface gravity defined in Eq.(\ref{surfaceG}). When $z =2$, it reduces to that obtained in \cite{ding}, and in the neutral case ($Q=0$) it further reduces to the one obtained in \cite{berglund2}.
It is clear that $T_{UH}^{z\ge 2}$ increases as $z$ becomes larger and larger, and finally reaches its maximum, $T_{UH}^{z = \infty}$, which is twice larger than $T_{UH}^{z = 2}$,
a limiting case that was also considered in \cite{JB13} without the presence of the electromagnetic field. It should be noted that the corresponding chemical potential always becomes unbounded at the universal horizons,
except for the three cases $z = 2,  3, \infty$, in which the chemical potential always vanishes.

{ As mentioned previously, to arrive at the above conclusions, we have implicitly assumed that each horizon, Killing or universal, is associated with a temperature. One cannot take this for granted, as the system
can be well   approximated as thermal only in a certain energy regime, but  not  in an equilibrium  state at all  \cite{Visser14}. This relies heavily on the full structure
of horizon thermodynamics,  and closely related to the underlaying theory at high energies. With this in mind,  we note that recently  the Horava theory was shown to be perturbatively renormalizable \cite{Bar}. In particular, its quantization in 2d spacetimes
reduces to that of a simple harmonic oscillator \cite{Li14}.  Therefore, it would be very interesting to study this important issue in a concrete  framework, the Horava theory of quantum gravity.}

 In addition, we have also studied the Smarr mass function formula, by assuming that: { (a) the entropy is proportional to the area of the universal horizon, and (b) } the first law of black hole thermodynamics  holds at the universal horizon.
 Together with the temperatures we have just obtained by  the   Hamilton-Jacobi method, these
 assumptions uniquely determines the Smarr mass, given by Eq.(\ref{BHmass}).
Applying it to the two particular black hole solutions of Eqs.(\ref{aether}) and (\ref{classB}), we have found that the corresponding  Smarr masses are
given, respectively, by Eqs.(\ref{BHmassA}) and (\ref{BHmassB}), { which are quite different from the well-known ones obtained in \cite{eling,garfinkle}. These differences imply that either the masses given in
\cite{eling,garfinkle} are incorrect, or  at least one of our above two assumptions must be modified. }

It would be extremely interesting to see if our results can be also obtained   when other methods are used \cite{KM,Carlip,unruh,MP09}.

\begin{acknowledgments}

We would like to thank  J. Bhattacharyya  Ted Jacobson Bao-Fei Li for valuable discussions and suggestions.  This work was done during C.D.'s visit to  Baylor University (BU) and he would like
to thank BU for hospitality.  A.W. is supported  in part by Ci\^{e}ncia Sem Fronteiras, No. A045/2013 CAPES, Brazil,  and NNSFC No.
11375153, China. C.D. is supported partly by the grants, NNSFC No. 11247013, China, Hunan Provincial NSFC No. 2015JJ2085, China, the State Scholarship Fund of China No. 201308430340,
 and the special fund of NNSFC No. 11447168, China. T.Z. is supported partly by  the Chinese grants,  NNSFC No. 11047008, No. 11105120, and No. 11205133, while X.W. is supported by BU
  through the physics graduate program.

\end{acknowledgments}

\appendix

\section{Fractional Derivatives }
\renewcommand{\theequation}{A.\arabic{equation}} \setcounter{equation}{0}

The fractional calculus is  a well-established branch of mathematics, and has been applied to physics widely. In the following, we just present some formulas
that have been used in this paper. For details, we refer readers to \cite{uchaikin,calcagni}.
First, the generalized Taylor series is given by,
\begin{eqnarray}
\lb{A.1}
f(x)=\sum^\infty_{j=0}\frac{(x-a)^{\alpha j}}{\Gamma(1+\alpha j)}{}_{a}f^{(\alpha j)}
(a),\quad0\leq\alpha<1,
\end{eqnarray}
where the left Caputo derivatives  is defined as
\begin{eqnarray}
\lb{A.2}
{}_af^{(\alpha)}(x)
&\equiv&\frac{1}{\Gamma(1-\alpha)}\int_a^x\frac{dx'}{(x-x')^{\alpha }}
\frac{df(x')}{dx'}, \; (x>a).\nonumber\\
\end{eqnarray}
Then, two useful fractional derivative formulas are \cite{calcagni}
\begin{eqnarray}
\lb{A.3}
&&\partial^{\alpha}(x-a)^\beta=\frac{\Gamma(\beta+1)}{\Gamma(\beta+1-\alpha)}
(x-a)^{\beta-\alpha},\nonumber\\&&
\partial^{\alpha}(a-x)^\beta =\frac{\Gamma(\beta+1)}{\Gamma(\beta+1-\alpha)}
(a-x)^{\beta-\alpha},\nonumber\\&&
~~~~~~~~~~~~~~~~ \quad\quad\quad\left(\beta\neq0,1,\cdots,n-1\right).
\end{eqnarray}

\section{Post-WKB method and Hawking Radiation at the Killing Horizon for the Case $z=2$} 
\renewcommand{\theequation}{B.\arabic{equation}} \setcounter{equation}{0}

From Section III B, one can see that the WKB treatment breaks down at the Killing horizon the cases  $z>1$. In this appendix, we shall use the Bremmer series method (it can be viewed as a post-WKB treatment) to study  the Hawking radiation  at the Killing horizon for the non-relativistic 
case $z=2$. We shall show that the corresponding temperature of Hawking radiation is also zero, the same as we obtained in Section III.B obtained by using the HJ method. This confirms  the validity of the HJ method for the Hawking radiation of non-relativistic particles. 
Note that the Post-WKB method is to use the WKB modes as a basis to  study solutions of second-order differential equations \cite{winitzki} or higher-order ones \cite{coutant16}. For details, we refer readers to \cite{winitzki,coutant16}.

To simplify the problem, following \cite{coutant16}, we shall ignore the $\theta,\varphi$ terms and study the problem   in the Pailev\'{e}-Gullstrand   coordinates. Then, the metric takes the form,
\begin{eqnarray}
ds^2=dt^2-(dr-vdt)^2,\;v\equiv-\sqrt{1-e(r)},
\end{eqnarray}
with the aether field
\begin{eqnarray}
u^a=(\cosh\theta,-(s\cdot\chi)),\;s^a=(-\sinh\theta,-(u\cdot\chi)),
\end{eqnarray}
where $\theta\equiv\theta(r)$ is a position-dependant boost angle relating to the four-vector ${\bf t}^a=\cosh\theta u^a-\sinh\theta s^a$, which defines the free-fall observer to aether frame \cite{JB13}. The propagating scalar field is described by  \cite{berglund2},
\begin{eqnarray}
\mathcal{L}=-\frac{1}{2}(\nabla\phi)^2-\frac{(\vec{\nabla}^2\phi)^2}{2k_0^2},
\end{eqnarray}
where $\vec{\nabla}^a=s^as^b\nabla_b$ is the projected (spatial) covariant derivation. Under this background, the dispersion relation is
\begin{eqnarray}\label{Omega}
(\omega-vk_r)^2=k_r^2+\frac{[-\omega\sinh\theta+k_r(u\cdot\chi)]^4}{k_0^2},
\end{eqnarray}
which is precisely the $z=2$ case of Eq.(\ref{ksquare}).
In the low frequency regime, we shall ignore the fourth-order term. Then, for $k<0$, Eq.(\ref{Omega}) can be rewritten as
\begin{eqnarray}\label{omega}
\omega=(1+v)k_r+\frac{(u\cdot\chi)^4}{2k_0^2}k_r^3.
\end{eqnarray}

The location of the turning point $r_{tp}$ can be obtained by solving $v_{gr}=\left(\partial k_r/\partial\omega\right)^{-1} =0$, where $v_{gr}$ is the group velocity \cite{countant12}. Hence, the momentum and the velocity at the turning point are
\bqn
-k_{tp}&=&\left[\frac{\omega k_0^2}{(u\cdot\chi)^4}\right]^{1/3},\nb\\
v(r_{tp})+1&=&-\frac{3}{2}\left(\frac{\omega
(u\cdot\chi)^2}{k_0}\right)^{2/3}.
\eqn
If $\omega$ is sufficiently small, the turning point is located in the near-horizon region, so
\bqn
v&=&-1+\kappa x,\;x=(r-r_{KH}),\nb\\
 \kappa&=&\left. \frac{\partial v}{\partial r}\right|_{r_{KH}},\;(u\cdot\chi)=u_h,
\end{eqnarray}
where $u_h\equiv (u\cdot\chi)|_{r_{KH}}$, and the turning point $x_{tp}$ is
\begin{eqnarray}
\kappa x_{tp}=-\frac{3}{2}\left(\frac{\omega
u_h^2}{k_0}\right)^{2/3}.
\end{eqnarray}
From Eq.(\ref{omega}) we can see that there is no turning point for $\omega > \omega_{max}$, where
\begin{eqnarray}
\omega_{max}=\frac{k_0}{u_h^2}\left[-\frac{2}{3}(v_{min}+1)\right]^{3/2},
\end{eqnarray}
where $v_{min}$ is the asymptotic value $v(x=-\infty)$.
By introducing the auxiliary functions
\begin{eqnarray}\label{Ux}
&&U^{\pm}_{\omega}(x)=\rho
\left[\frac{\omega}{\rho}\pm\sqrt{\frac{8\kappa^3x^3}{27}+\frac{\omega^2}{\rho^2}
}\right]^{1/3},
\end{eqnarray}
where $\rho\equiv k_0/u_h^2$, we find that the three roots of (\ref{omega}) are combinations of $U^{\pm}_{\omega}(x)$.

Now we are in a position to derive the local scattering coefficients $\alpha$ and $\beta$ \cite{coutant16}. The modulus of $\alpha$ is
\begin{eqnarray}
|\alpha|^2\sim|C|^2\exp\left[-2\text{Im}S\right],\;S\equiv\int_{x_0}^{x_{tp}}
\big(k_u(x')-k_+(x')\big)dx',\nonumber
\end{eqnarray}
where $C$ is a constant prefactor, $x_0$ is a real reference point, which can be chosen anywhere, and $k_u,k_+$ are solutions of Eq.(\ref{omega}).
By using Eq.(\ref{Ux}), we find that the difference of these two solutions is
\begin{eqnarray}
k_u(x')-k_+(x')&=&\sqrt{3}i[U^+-U^-]
\nonumber\\&\approx& i\rho\sqrt{2\kappa x}.
\end{eqnarray}
So, we find that 
\begin{eqnarray}
S&=&\int_{x_0}^{x_{tp}}
i\rho\sqrt{2\kappa x'}dx'=\frac{\rho}{\kappa}\int^{\frac{3}{2}
(\frac{\omega}{\rho})^{2/3}}_{t_0}\sqrt{2t}dt\nonumber\\
&=&\frac{\sqrt{2}\omega}{\kappa}-\frac{2\sqrt{2}\rho}{3\kappa}t_0^{3/2},
\end{eqnarray}
where $t=-\kappa x'$. It is real and has no contribution to scattering coefficient $|\alpha|^2$. Therefore, there is no Hawking radiation at the Killing horizon for the case $z = 2$,  which is the same as what we obtained in Section III.B.


\begin{thebibliography}{99}
\bibitem{NoteA} It can be shown that
the khronon field $\phi$ in the model studied in \cite{michel}
is given by
\bqn
 \lb{KH2}
 \phi(v, r) = \begin{cases}
 v - r - M\ln\left|r - M\right| + \phi_{+}, & v > v_0,\cr
 v - r + \phi_{-}, & v < v_0,\cr
 \end{cases}
 \nb
 \eqn
 where $\phi_{\pm}$ are two integration constants, $v = v_0$ is the location of the collapsing null shell, and ($v, r$) denote  the   Eddington-Finklestein coordinates [cf. Eq.(\ref{metric})].
 For more detail, see \cite{michel}.
\end{thebibliography}
\end{document}